\RequirePackage{fix-cm}
\documentclass[smallextended,english]{svjour3}       
\smartqed  
\usepackage{graphicx}
\usepackage[T1]{fontenc}
\usepackage[latin9]{inputenc}
\usepackage{amsmath}
\usepackage{graphicx}
\usepackage{amssymb}
\usepackage{esint}

\makeatletter
\@ifundefined{textcolor}{}
{%
 \definecolor{BLACK}{gray}{0}
 \definecolor{WHITE}{gray}{1}
 \definecolor{RED}{rgb}{1,0,0}
 \definecolor{GREEN}{rgb}{0,1,0}
 \definecolor{BLUE}{rgb}{0,0,1}
 \definecolor{CYAN}{cmyk}{1,0,0,0}
 \definecolor{MAGENTA}{cmyk}{0,1,0,0}
 \definecolor{YELLOW}{cmyk}{0,0,1,0}
 }

\global\long\def\im{\text{Im}}

\newcommand{\be}{\begin{eqnarray}}\newcommand{\ee}{\end{eqnarray}}\def\beq{\begin{equation}}\def\eeq{\end{equation}}

\makeatother

\usepackage{babel}

\makeatother

\begin{document}
\title{Quantifying fluctuations from the tunnelling differential conductance}
\author{Antonio M. Garc\'{\i}a-Garc\'{\i}a, Pedro Ribeiro}
\institute{Antonio M. Garc\'{\i}a-Garc\'{\i}a \at
              Cavendish Laboratory, JJ Thomson Avenue,
Cambridge, CB3 0HE, UK,
              \email{amg73@cam.ac.uk}           
           \and Pedro Ribeiro
            \at Max-Planck-Institut fur Physik komplexer Systeme, Nothnitzer Str. 38, 01187 Dresden, Germany,
              \email{pribeiro@pks.mpg.de}}
\date{Received: date / Accepted: date}
\maketitle
\begin{abstract}
Dynes ansatz \cite{dynes} is a broadly utilized procedure to extract the superconducting energy gap from the tunnelling differential conductance, a quantity that is measured by scanning tunnelling microscopy (STM) techniques. 
In this letter we investigate the limit of applicability of this ansatz in nano-scale superconductors and propose a generalization that permits to study thermal and quantum fluctuations in STM experiments.
\end{abstract}
\section{ Introduction}
Scanning tunneling microscopy techniques (STM) have revolutionized \cite{stm1} the field of experimental superconductivity for both conventional and high temperature superconductors. A typical observable in these experiments is the normalized tunnelling differential conductance,
\begin{eqnarray}
\frac{dI}{dV} & = & \frac{1}{4T}\int_{-\infty}^{\infty}d\omega\, N_{s}(\omega,\Delta)\left[\frac{1}{\cosh^{2}\left(\frac{\omega+eV}{2k_{B}T}\right)}\right]\label{didv}\end{eqnarray} 
where $N_s$ is the local density of states, $T$ is temperature, $V$ is the gate voltage.
In order to make contact with properties of the superconductor, such as the superconducting energy 
 gap $\Delta$, is required to fit the experimental data with a theoretical expression of $dI/dV$ that includes $\Delta$ as a fitting parameter. In principle this is not an easy task as for many materials the theoretical 
 local density of states $N_s(\omega,\Delta)$ 
 is not exactly known or its calculation involves the use of the rather cumbersome Eliashberg theory of superconductivity.
However Dynes noted \cite{dynes} that for some metals the energy gap $\Delta \equiv \Delta_D$ obtained by a simple fitting function,
 \begin{eqnarray}
N_{D}(\omega) & = & \im\left(\frac{\omega+i\Gamma}{\sqrt{\Delta_D^{2}-\left(\omega+i\Gamma_D\right)^{2}}}\right) \label{DDOS}\end{eqnarray}
is in excellent agreement with the one obtained by other experimental techniques. $\Gamma_D(T)$ in (\ref{DDOS}) is other fitting parameter that describes different sources of decoherence.
 In light of this success this procedure is now employed to obtain the energy gap in STM experiments involving a broad range of conventional and high $T_c$ superconductors. In case of a material with a non s-wave order parameter symmetry the Dynes ansatz is modified to account for the expected momentum dependence of $\Delta_D$.\\  
 In this letter we investigate the limit of applicability of (\ref{DDOS}) in nano-scale superconductors where deviations from mean-field predictions are well documented \cite{richardson,nmat}. These were recently used in \cite{nmat1} to analyze STM data from Pb superconducting nano particles. We focus on  sizes $L < 10$nm for which deviations from mean-field predictions start to be relevant. First we introduce a model capable to describing thermal fluctuations and provide theoretical expressions for the gap and $dI/dV$. Then we fit this $dI/dV$ by Dynes ansatz (\ref{DDOS}),(\ref{didv}) and compare the resulting gap with (\ref{SPAgap}).Finally we propose a generalization of Dynes ansatz that broadens the information obtained from STM experiments. 
 \section{Theoretical description of thermal fluctuations in zero dimensional superconductors}
In order to test Dynes procedure in the nanoscale we study the reduced BCS Hamiltonian by path integral techniques 
in the static path approximation (SPA) \cite{static} which provide a good description of thermal fluctuations in 
 the zero dimensional limit corresponding to grain sizes much smaller than
the superconducting coherence length. Here we only present the main results and refer to \cite{static,nmat1} for details. 
The partition function $Z$ is given by,  
\begin{eqnarray}
\frac{Z}{Z_{0}} & = & \int d\left|\Delta\right|\,\left|\Delta\right|\, e^{-\beta\delta^{-1}\mathcal{A}(\left|\Delta\right|)}\end{eqnarray}
with \begin{eqnarray}
\mathcal{A}\left(\left|\Delta\right|\right) & = & \lambda^{-1}\left|\Delta\right|^{2}-\int_{-E_{D}}^{E_{D}}d\varepsilon\,\varrho\left(\varepsilon\right)\left[\left(\xi-\left|\varepsilon\right|\right)+\frac{2}{\beta}\log\left(\frac{e^{-\beta\xi}+1}{e^{-\beta\left|\varepsilon\right|}+1}\right)\right] \end{eqnarray}
where $\xi=\sqrt{\varepsilon^{2}+\left|\Delta\right|^{2}}$, $\beta  = 1/T$ and $\varrho\left(\varepsilon\right)=\sum_{\alpha}\delta\left(\varepsilon-\varepsilon_{\alpha}\right)$
is the one particle density of states, $Z_{0}$ is the parition function for non-interacting electrons, $\lambda$ is the dimensionless coupling constant, $E_D$ is the Debye energy and $\delta$ is the mean level spacing. From the explicit knowledge of $Z$, the normalized density of states (DOS) $N_{s}(\omega)$ is,  
\begin{eqnarray}
N_{s}(\omega) & = & \left(\frac{Z}{Z_{0}}\right)^{-1}\int d\left|\Delta\right|\,\left|\Delta\right|e^{-\beta\delta^{-1}\mathcal{A}(\left|\Delta\right|)}\int_{-E_{D}}^{E_{D}}d\varepsilon\,\varrho \left(\varepsilon\right)\im\left[\frac{\omega + i \Gamma_S}{\sqrt{\Delta^2-(\omega - i \Gamma_S)^2}} \right].\label{SPADOS}\end{eqnarray}
where $\Gamma_S(T)$ is a phenomenological parameter that describes all decoherence mechanisms except thermal fluctuations.
The superconducting gap $\bar \Delta$ in this formalism is given by, \begin{eqnarray}
{\bar \Delta}^{2} & = & \left(\frac{Z}{Z_{0}}\right)^{-1}\int d\left|\Delta\right|e^{-\beta\delta^{-1}\mathcal{A}(\left|\Delta\right|)}\left|\Delta\right|^{3}\left[\lambda\int_{-E_{D}}^{E_{D}}d\varepsilon\,\varrho\left(\varepsilon\right)\frac{\tanh\left(\frac{\beta\xi}{2}\right)}{2\xi}\right]^{2}\label{SPAgap}.\end{eqnarray}
We note that $\bar \Delta$ leads to the bulk gap $\Delta_0 = \dfrac{E_D}{\sinh{1/\lambda}}$ for $L \to \infty$ however finite size effects in $\bar \Delta$ might differ slightly from those in the spectral gap \cite{richardson}. 
\section{Justification of Dynes ansatz in the nanoscale region}
\begin{figure}
\includegraphics[width=1\columnwidth]{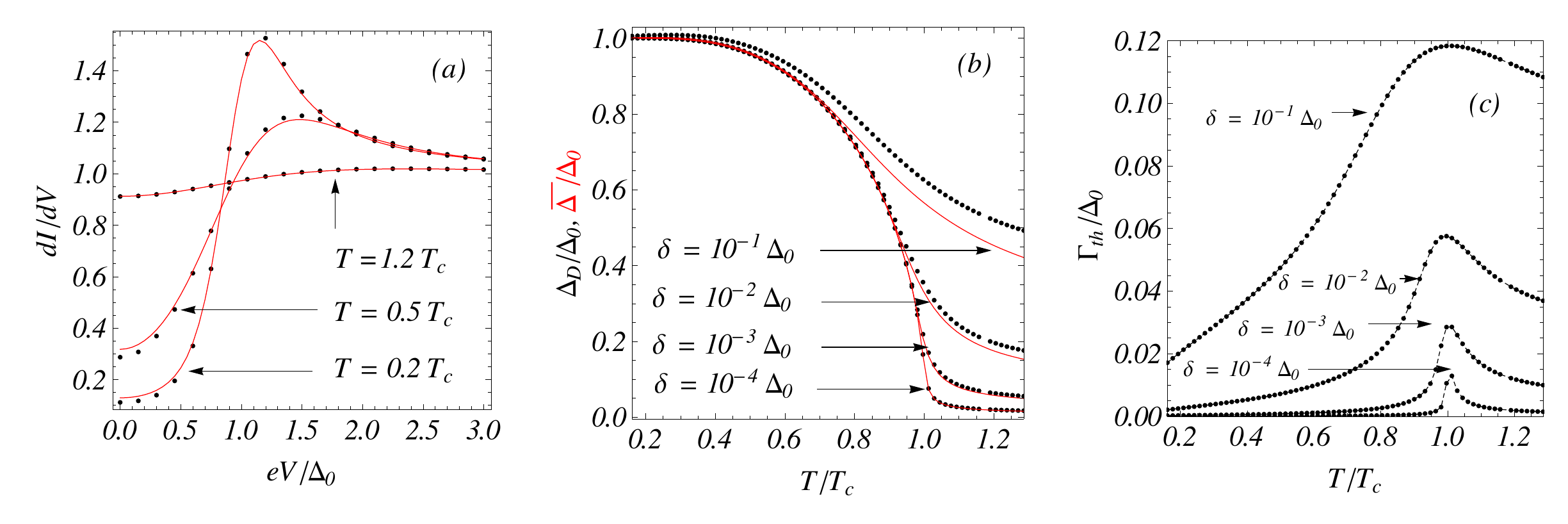}\caption{\label{Dynes_SPA}(a) $dI/dV$ (\ref{didv}) as a function of $eV$ for 
$\delta = 0.1\Delta_0$, $\Gamma = 0.1\Delta_0$, $E_D = 9\Delta_0$, $\lambda=0.34$. Red lines stand for the theoretical $dI/dV$ from (\ref{DDOS}) and dots correspond to Dynes fitting (\ref{SPADOS}). We assume $\rho(\epsilon) \approx 1/\delta$,
 (b) the order parameter for the same parameters as in (a).
Red lines ${\bar \Delta(T)}$ (\ref{SPAgap}) correspond to the SPA prediction 
and black dots  $\Delta_{D}(T)$ stand for Dynes fitting, (c) $\Gamma_{\text{th}}(T)$
for the same parameters as in (b) (see text for more details). }
\end{figure}

We now test Dynes ansatz by comparing $\Delta_D(T)$ (\ref{DDOS}) with ${\bar \Delta}(T)$ (\ref{SPAgap}) where the former is obtained by fitting 
the theoretical $dI/dV$ (obtained from (\ref{SPADOS})) with the Dynes $dI/dV$ coming from (\ref{DDOS}).
Dynes ansatz is presumably applicable provided that the low energy
excitations are well defined quasiparticles - Fermi liquid theory holds - and a gap do 
exist in the spectrum such that the quasiparticle dispersion relation is $E = \sqrt{\epsilon^2+\Delta^2}$ even if $\Delta$ is not the BCS prediction. 
We show in Fig. \ref{Dynes_SPA}a that Dynes expression fits very well the theoretical $dI/dV$ and in Fig. \ref{Dynes_SPA}b that the energy gap $\Delta_D(T)$, obtained from (\ref{DDOS}), is in excellent
agreement with the theoretical prediction (\ref{SPAgap}) that includes thermal fluctuations. As $\delta$ increases a small difference is observed. It is therefore likely that, as fluctuations become dominant, (\ref{DDOS}) is less accurate. 
It is therefore justified to employ Dynes ansatz to fit the experimental $dI/dV$ in the nanoscale region provided that fluctuations are not dominant $\delta \ll {\rm max}(\Delta_0,T_c)$. For larger fluctuations the generalized expressions (\ref{SPADOS}), (\ref{SPAgap}) provides a suitable generalization of Dynes expression in the nanoscale where now the fitting parameters are $\lambda$ and $\Gamma_S$ but $\Gamma_S \neq \Gamma_D$ since $\Gamma_D$ does not include the effect of thermal fluctuations. Therefore this generalization of the Dynes ansatz allows to single out the effect of fluctuations from other forms of decoherence.  
\section{Quantifying fluctuations in STM experiments}
The analysis of experimental STM data by the SPA formalism (\ref{SPADOS}), (\ref{SPAgap}) developed above has clear advantages over (\ref{DDOS}) even for sizes in which Dynes ansatz describes well the energy gap.
We focus on the study of $\Gamma(T)$ which includes effects leading to a finite quasiparticle life-time and thermal fluctuations as well. 
We note that $\Gamma=\Gamma_{D}$ in (\ref{DDOS}) englobes two contributions: $\Gamma_{D}=\tau^{-1}+\Gamma_{\text{th}}$ where 
the former accounts for the quasiparticle life time and 
the latter is related to thermal
fluctuations arising due to finite size effects. It is therefore expected that $\Gamma_{\text{th}}$
has a peak around the mean-field critical temperature with a typical width that describes the interval of temperatures where fluctuations are important. 
A important drawback of Dynes approach is that it is not possible to disentangle these two contributions. 
The situation is different if the SPA formalism (\ref{SPADOS}),(\ref{SPAgap}) is employed. Here thermal fluctuations are included so that $\Gamma_S \equiv 1/\tau$ {\it only} accounts for the finite quasiparticle lifetime. This is an important difference. 
 One can quantitatively estimate the role of thermal fluctuations by
the quantity $\Gamma_{\text{th}}=\Gamma_{D}-\Gamma_S$ which can be obtained by fitting the experimental data by both Dynes ansatz (\ref{DDOS}) and the SPA expression (\ref{SPADOS}). This quantity has a maximum (see Fig. \ref{Dynes_SPA}c) near the critical temperature where thermal fluctuations are more important. 
Therefore it can be used to estimate: a)the importance of thermal fluctuations with respect to other sources of decoherence, b) the (would-be) critical temperature and c) the region around $T_c$ where fluctuations are relevant. Similar findings are applicable to $T \ll T_c$ provided that quantum fluctuations are also included in the theoretical formalism. In summary, the use of a theoretical framework that takes into account deviations from mean-field to analyse experimental data expands substantially the information that can be obtained from STM experiments.   

\end{document}